\documentclass[aps,prd,amsfonts,amsmath,nofootinbib,showpacs,preprint,tightenlines,longbibliography]{revtex4-1} 
\usepackage{epsf}

\def\be{\begin{equation}}
\def\ee{\end{equation}}
\def\bea{\begin{eqnarray}}
\def\eea{\end{eqnarray}}
\def\bml{\begin{subequations}}

\def\elea{\end{eqnarray}\end{subequations}}

\begin{document}

\title{Direct determination of cosmic string loop density from simulations}

\author{Jose J. Blanco-Pillado}
\email{josejuan.blanco@ehu.es}
\affiliation{Department of Theoretical Physics, UPV/EHU,\\48080, Bilbao, Spain}
\affiliation{IKERBASQUE, Basque Foundation for Science, 48011, Bilbao, Spain}

\author{Ken D. Olum}
\email{kdo@cosmos.phy.tufts.edu}
\affiliation{Institute of Cosmology, Department of Physics and Astronomy,\\Tufts University, Medford, MA 02155, USA}

\begin{abstract}
We determine the distribution of cosmic string loops directly from
simulations, rather than determining the loop production function and
inferring the loop distribution from that.  For a wide range of loop
lengths, the results agree well with a power law exponent $-2.5$ in
the radiation era and $-2$ in the matter era, the universal result for
any loop production function that does not diverge at small scales.
Our results extend those of Ringeval, Sakellariadou, and Bouchet: we
are able to run for 15 times longer in conformal time and simulate a
volume 300--2400 times larger.  At the times they reached, our
simulation is in general agreement with the more negative exponents
they found, $-2.6$ and $-2.4$.  However, our simulations show that
this was a transient regime; at later times the exponents decline to
the values above.  This provides further evidence against models with
a rapid divergence of the loop density at small scales, such as
``model 3'' used to analyze LIGO data and predict LISA sensitivity.

\end{abstract}

\maketitle

\section{Introduction}

Cosmic strings are long, thin objects which may form by the Kibble
mechanism during a symmetry-breaking transition in the early universe \cite{Kibble:1976sj,Vilenkin:2000jqa},
or they may be the strings of string theory (or one-dimensional
D-branes) left over from a string-theory inflationary scenario \cite{Dvali:2003zj,Copeland:2003bj}.  In
the simplest scenarios, which we consider here, strings have no ends,
but exist in a ``network'' made up of infinite strings and
a distribution of loops of all sizes. 

Intercommutation breaks off loops from infinite strings or from larger
loops.  Over time, energy flows from infinite strings into loops,
and is eventually radiated in gravitational waves.  (We will consider
only the case where there is no other channel through which loops may
lose their energy.)  This flow of energy prevents strings from coming
to dominate the universe as monopoles would.  Instead the string
network scales, with all linear measures of string properties being
proportional to the age of the universe $t$.  The infinite string
density scales as radiation in radiation era and as matter in the
matter era.

Many potentially observable signals of a cosmic string network depend
on the loop distribution.  In particular, the best way of discovering
a cosmic string network is to observe the gravitational waves coming
from loops
\cite{Damour:2004kw,Siemens:2006vk,Sanidas:2012ee,Binetruy:2012ze,
  Sousa:2016ggw,Kuroyanagi:2012wm,Blanco-Pillado:2017oxo,Abbott:2017mem,Auclair:2019wcv}. The
non-observation of such signals, by pulsar timing experiments,
currently provides the strongest bounds on the energy scale of the
string network \cite{Blanco-Pillado:2017rnf}. Thus it is very
important to determine the distribution of cosmic string loops.

Two methods have been used to determine the loop distribution.  The
simplest is just to examine the distribution of loops existing at
several times during the simulation.  Then exhibit the distribution in
scaling units, so that in the scaling regime, it would not change with
time.  Identify the section of the distribution that seems to be
unchanging to find the scaling distribution.  This technique was used
early on by Albrecht and Turok \cite{Albrecht:1989mk}, and then by
Ringeval, Sakellariadou and Bouchet~\cite{Ringeval:2005kr}, who found
that the loop number density scaled as loop length to the power
-2.60 in the radiation era and
-2.41 in the matter era.\footnote{These are the central values.  The
  range of possibilities is discussed below.}

The second technique is to study the rate at which loops of different
sizes are produced.  By showing this in scaling units, determine the
scaling loop production function.  Then use cosmological kinematics to
go from the loop production function to the loop distribution.  This
technique also was used by Albrecht and Turok \cite{Albrecht:1989mk}
and then by Blanco-Pillado, Olum, Shlaer, Vanchurin, and
Vilenkin~\cite{Vanchurin:2005pa,Olum:2006ix,BlancoPillado:2011dq,Blanco-Pillado:2013qja}
They found exponents -2.5 (radiation) and -2.0 (matter).
(Reference~\cite{BlancoPillado:2011dq} also plotted the the loop
distribution directly and found agreement with these exponents.)

The exponents -2.5 and -2.0 are universal in the sense that if loop
production is limited to a certain range of sizes, the loop distribution
will have these exponents for all smaller sizes.  In fact, the only
way for a simulation to yield more negative exponents is for the loop
production function to diverge at the same rate, which is forbidden by
energy conservation~\cite{Blanco-Pillado:2019vcs}.

In this paper, we use the the same simulation code that we have used
in earlier papers \cite{BlancoPillado:2011dq,Blanco-Pillado:2013qja}
but determine the loop distribution directly, rather than working
through the loop  distribution function.  This allows a direct
comparison with Ref.~\cite{Ringeval:2005kr}.

To find the loop distribution relevant to cosmology we must modify the
distribution from simulations to take account of gravitational effects
on loops and long strings.  Loops lose their energy to gravitational
waves, shrinking and eventually vanishing, and long strings become
smoother, affecting the loop emission process.  However, we will
discuss gravitational processes only very briefly here, and instead
concentrate on the first step: determining the loop distribution from
simulations.

\section{Loop distribution}

Scaling means that all linear measures of a string network evolve as a
fixed multiple of the age of the universe $t$ or the distance to the
horizon $d_h$.  All other measures scale according to their dimension,
and distributions are invariant when they are written in terms of
scaling quantities, as we do below.

In a simulation we start with some initial conditions, which are not
exactly the right ones for a scaling network.  As time passes we expect many
properties of the simulated network, such as the long string density,
to approach the scaling regime.  However, we cannot expect complete
scaling of the network in a simulation, because there is no way for
energy to be lost from the network.  Once a loop is much smaller than
the horizon, its physical size stays the same, so its comoving size
shrinks.  As a result, the loop distribution for large loops should
scale, but there is always a nonscaling distribution of small loops
and the total amount of string in the universe does not scale.  The
dividing comoving size between the scaling and the nonscaling parts
of the loop distribution decreases with time.  Gravitational effects
would remedy these problems and produce a scaling distribution at all
sizes, but we are not considering those here.
        
We will describe loops sizes with a scaling variable $x = l/t$, where
$l$ is the invariant length of the loop, meaning that the energy of
the loop is $\mu l$, where $\mu$ is the string tension.  This is the
same definition as Ref.~\cite{Blanco-Pillado:2019vcs}, but 
Refs.~\cite{Ringeval:2005kr,BlancoPillado:2011dq,Blanco-Pillado:2013qja}
used $l/d_h$ as their measure of scaling length.  In a power law
cosmology where the scale factor $a$ goes as $t^\nu$, the horizon
distance is $t/(1-\nu)$, so to convert $l/d_h$ into $l/t$ we must
divide by $(1-\nu)$.

We will describe the loop distribution by a function $n(x)$ where
$n(x) dx$ is the average number of loops in volume $t^3$ with sizes
between $x$ and $x+dx$.  To convert from loop densities based on
$l/d_h$ to the present convention we multiply by $(1-\nu)^4$.  For a
table of different notations used in some recent papers, see the
appendix of Ref.~\cite{Blanco-Pillado:2019vcs}.

In order to specify a loop distribution, one must clearly define
what constitutes a loop.  In our work, loops are only counted as loops
when they are in non-self-intersecting trajectories which will not
rejoin other loops or long strings.  We cannot apply the latter test
perfectly, because we only wait a finite amount of time (between 2 and
3 oscillations of the loop) to see if it will rejoin.  However, since
loops retain their physical size while the rest of the network is
rapidly diluted in physical units, it is quite unlikely for a loop to
rejoin if it has not done so in the first few oscillations.  It is
especially unlikely if the loop is very small, so there is certainly no
significant impact on the small-$x$ form of the loop distribution.

Ref.~\cite{Ringeval:2005kr} simply counts any loop of string smaller
than the horizon as a loop.  Loops near the horizon size are very
likely to rejoin, so we would not usually count them as loops.  The
result is that our loop distributions fall at large sizes relative to
the extrapolation of the small-size behavior, whereas theirs increase.

\section{Results}

We simulated cosmic string networks starting from
Vachaspati-Vilenkin~\cite{Vachaspati:1984dz} initial conditions and
directly accumulated the loop distributions at a set of
logarithmically spaced times.  In the radiation era, we ran until
conformal time 1500, and in the matter era until conformal time 750,
in units of the initial Vachaspati-Vilenkin cell size.  These are both
15 times larger than the conformal times reached in
Ref.~\cite{Ringeval:2005kr}.  The comoving volume of space simulated
was larger by a factor of about 2400 in the radiation era and 300 in
the matter era than that used by Ref.~\cite{Ringeval:2005kr}.  For
details about our simulations see Ref.~\cite{BlancoPillado:2011dq}.

We will compare our results with those of
Refs.~\cite{Ringeval:2005kr,Blanco-Pillado:2013qja}. Both references
claimed that the scaling loop distribution at small $x$ was a power
law, which we will write $n(x) = Cx^{-\beta}$. In that notation,
Ref.~\cite{Ringeval:2005kr} found
\begin{align}\label{eqn:C1r}
C &= 0.08\pm0.05 &\beta &= 2.60^{+0.21}_{-0.15}&\text{(radiation)}\\
\label{eqn:C1m}
C &= (1.5\pm0.5)\times 10^{-2} &\beta &= 2.41^{+0.08}_{-0.07}&\text{(matter)}\,.
\end{align}
Meanwhile, in Ref.~\cite{Blanco-Pillado:2013qja} we found
\begin{align}
C &= 0.18 & \beta &= 2.5&\text{(radiation)}\\
\label{eqn:C4}
C &= 0.27 & \beta &= 2.0&\text{(matter).}
\end{align}
by extrapolating from the loop production function.

\subsection{Radiation era}

\begin{figure}
\centering
\epsfxsize=6in\epsfbox{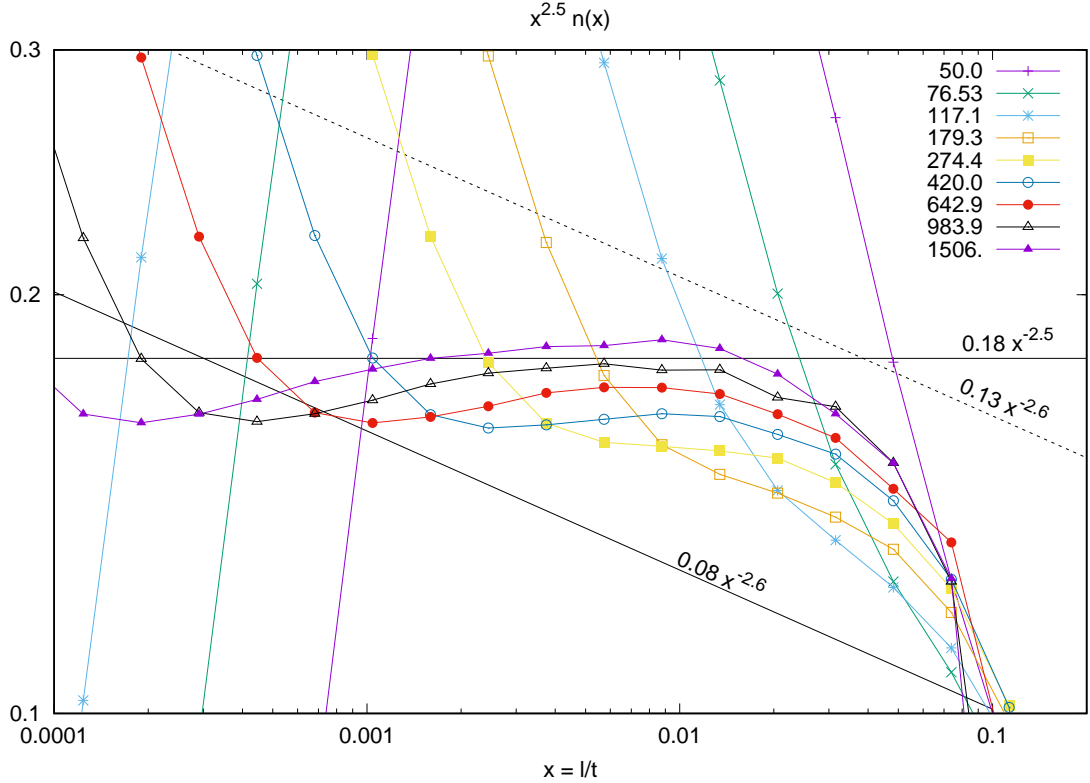}
\caption{Loop densities in the radiation era at the conformal times
  listed, computed from five simulations of size 1500.  The scaling
  loop density has been multiplied by $x^{2.5}$ to make small
  differences easier to see.  The vertical scale ranges only over of a
  factor of 3 in loop density.  Compare with more than 7 orders of
  magnitude in Fig.~3 of Ref.~\cite{Ringeval:2005kr} and 14 orders of
  magnitude in Fig.~9 of
  Ref.~\cite{BlancoPillado:2011dq}.}\label{fig:radiation}
\end{figure}

Figure~\ref{fig:radiation}
shows our results for the radiation era.  The colored lines with data
points are the $n(x)$ found directly from our simulations.  The
horizontal black line is $n(x) = 0.18 x^{-2.5}$ as predicted by
Eq.~(63) of Ref.~\cite{Blanco-Pillado:2013qja}.  At the last time
shown, the actual loop density is close to this line for loops whose
lengths are in the range $10^{-3}$--$10^{-2}$.  At larger sizes, we
would not expect to this approximation to apply, because loops there
have not yet been created.  At smaller sizes, the loop density drops
off slowly and then rises rapidly.  The drop-off occurs because it
takes time to approach a scaling regime.  Scaling loops with
$x=10^{-4}$ at conformal time 1500, for example, were primarily
produced around time 150, when the network was much further from
scaling and the production rate of large loops was less.  The rapid
rise at very small $x$ is an artifact of the initial conditions.

Even by time 1500, we have not reached a scaling loop density.  The
loop density takes longer to scale than the loop production function,
because it is sensitive to loops produced in earlier eras further from
scaling.  For this reason we do not recommend using loop distributions
to predict astrophysical signals.  It is more accurate to extrapolate
from the loop production function, as we did in
Ref.~\cite{Blanco-Pillado:2013qja}.  We exhibit the directly
determined loop distributions here only for comparison with prior
work.

In Fig.~\ref{fig:radiation} we see that the loop distribution has the
same basic form for the last 3 or 4 timesteps shown.  Before that
there is a transient regime in which the loop distribution has a
relatively straight segment rising more rapidly toward smaller loop
sizes.  This transient regime agrees reasonably well with the results
of Ref.~\cite{Ringeval:2005kr}.  That paper was able to reach only
conformal time 100.  The slanted black line in
Fig.~\ref{fig:radiation} shows the central value of the extrapolated
loop density determined there at the end of the run.  This line runs
mostly parallel to our result for time 179.3.  The difference in
normalization is of no consequence, because the error bars given in
Ref.~\cite{Ringeval:2005kr} extend upward to the dashed line.  The
loop density was far from scaling at that time.  Since we did not use
the same initial conditions as Ref.~\cite{Ringeval:2005kr}, we should
not expect close agreement before scaling is reached.  Nevertheless,
one can see the general effect that times around 100 there is a
steeper slope, but this is a transient that is replaced at later times by
a curve with exponent around $-2.5$, falling off faster toward smaller
sizes for the reason explained above.

A power law with $\beta=2.5$ lies within the error bars found by
Ref.~\cite{Ringeval:2005kr}.  Thus our results here are consistent
with their results.  However, later
work~\cite{Lorenz:2010sm,Abbott:2017mem,Auclair:2019wcv} based on
Ref.~\cite{Ringeval:2005kr} used extrapolations applicable only for
$\beta>2.5$.  Such extrapolations are not justified by the larger
simulations reported here.

\subsection{Matter era}

\begin{figure}
\centering
\epsfxsize=6in\epsfbox{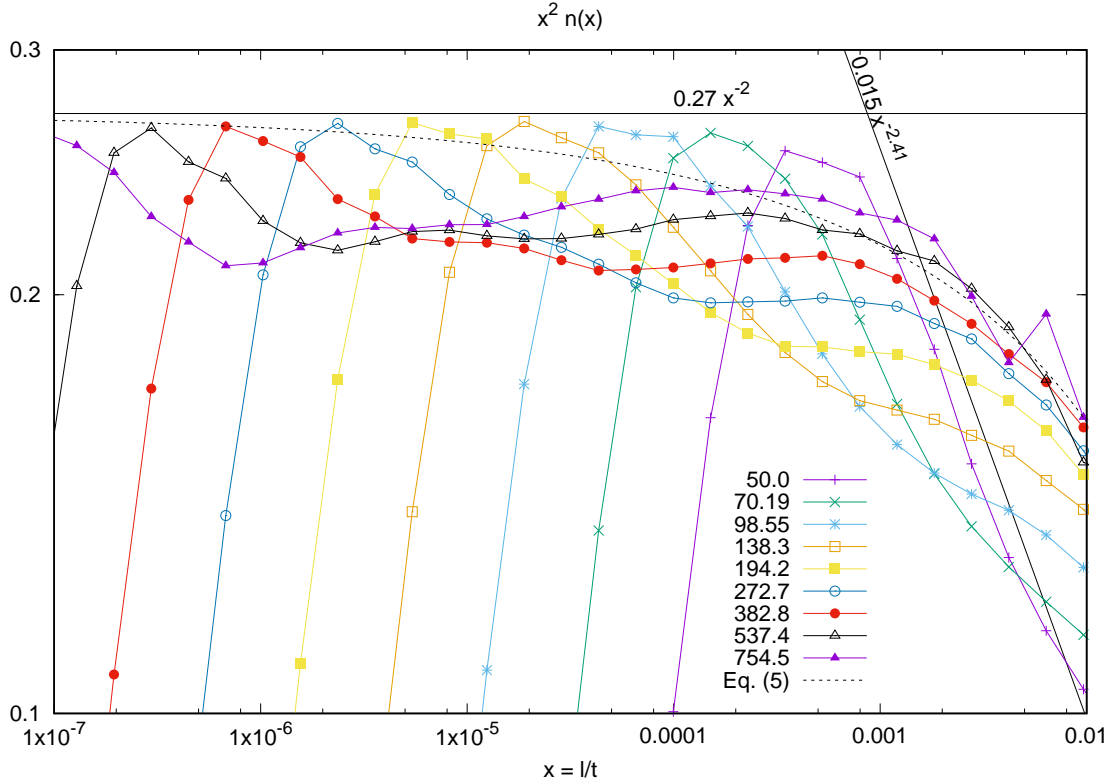}
\caption{Loop densities in the matter era at the conformal times
  listed, computed from two simulations of size 750.  The scaling loop
  density has been multiplied by $x^2$ to make small differences
  easier to see.  The vertical scale ranges only over of a factor of 3
  in loop density, compared with 6 orders of magnitude in Fig.~3 of
  Ref.~\cite{Ringeval:2005kr} and 11 orders of magnitude in Fig.~6 of
  Ref.~\cite{BlancoPillado:2011dq}.}\label{fig:matter}
\end{figure}

Figure~\ref{fig:matter}
shows our results for the matter era.  The situation is somewhat more
complicated here because in the matter era the loop production is
spread out over a wider range of loop sizes.  The horizontal black
line is $n(x) = 0.27 x^{-2}$, which is the asymptotic value for small
loops predicted by Eq.~(65) of Ref.~\cite{Blanco-Pillado:2013qja}.
The curved, dashed black line is
\be\label{eqn:matterfit}
n(x) = 0.27 x^{-2}- 0.45x^{-2.31}\,,
\ee
the entire prediction of the same equation without gravitational
effects.  We see that the directly computed loop density matches
reasonably well to this line above about $x=3\times 10^{-5}$.  For
smaller loops there is a decline because loop production was not
scaling at early times, and for much smaller loops a nonscaling rise
associated with the initial conditions.

The sharply angled black line is the prediction of
Ref.~\cite{Ringeval:2005kr} based on simulation data through conformal
time 50 in units of the initial correlation length.  Our simulations
at time 50 agree very well with this distribution, but that is a
transient regime, and the results at later times are of a different
nature entirely.  Taking account of the error bars in
Eq.~(\ref{eqn:C1m}) does not produce much better agreement with the
late-time data in Fig.~\ref{fig:matter}.

\subsection{Smoother initial conditions}

In the simulations shown above the initial conditions consist of
strings that go in straight segments between the centers of the
Vachaspati-Vilenkin~\cite{Vachaspati:1984dz} plaquettes.  One might be
concerned that these strings are unnatural because they have
significant kinks as they pass through the plaquette centers.  To make
sure that our results are not affected by such artifacts, we did some
additional runs considering the string to be a quarter circle as it
goes between two adjacent faces of a Vachaspati-Vilenkin cell.  We
then represented each quarter circle by 20 linear pieces.  (If the
string connects two opposite faces it is just a straight line, so
there is no need to divide it.)  Since these simulations have much
more data than the ones with a single, straight segment between cube
faces, we used we a smaller box and ran only until conformal time 420.

This division of circular segments into 20 smaller pieces corresponds
roughly to the 20 ``points per correlation length'' used by
Ref.~\cite{Ringeval:2005kr}, but the ideas are not entirely the same.
In our case, this choice affects the initial conditions only, but in
Ref.~\cite{Ringeval:2005kr}, the point spacing also affects the
evolution of the network.  Nevertheless, these runs may be more
directly comparable to those of Ref.~\cite{Ringeval:2005kr}.

\begin{figure}
\centering
\epsfxsize=6in\epsfbox{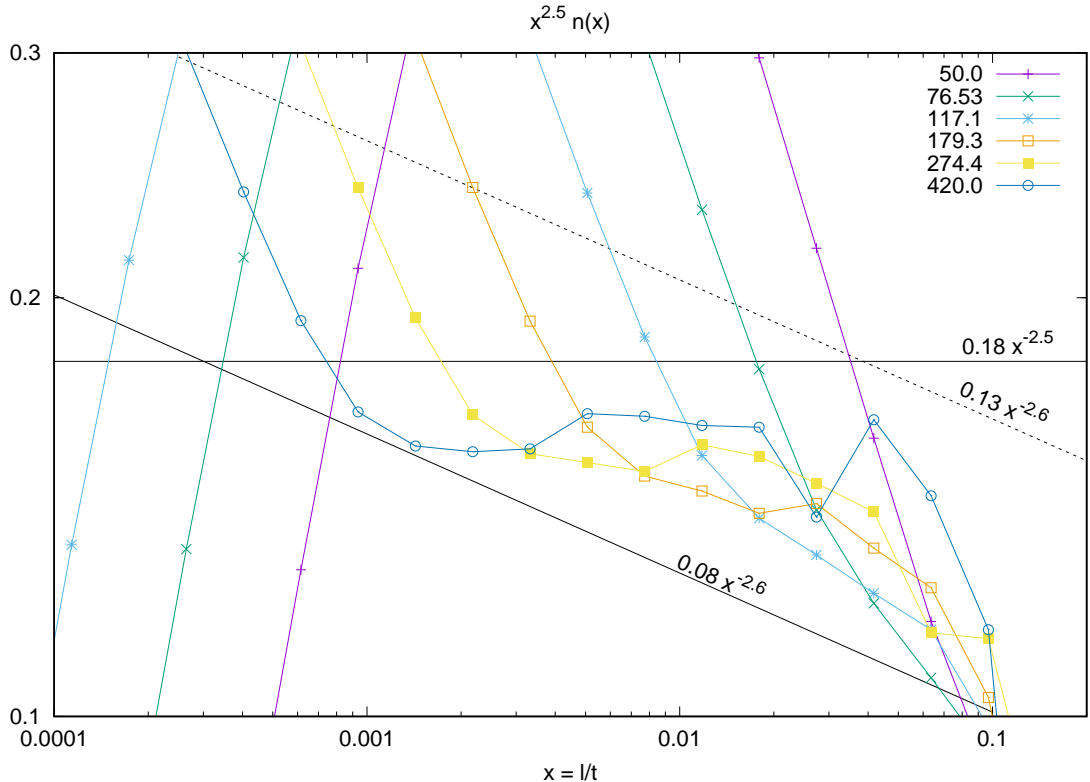}
\caption{Loop densities in the radiation era computed from five
  simulations of size 400 with curved segments in the initial
  conditions represented in 20 pieces.  The 6 times shown here are the
  same as the first 6 times in
  Fig.~\ref{fig:radiation}.}\label{fig:radiation-20}
\end{figure}

The resulting loop density is shown in
Fig.~\ref{fig:radiation-20}.
The results in this case are generally similar to those of
Fig.~\ref{fig:radiation}, showing that the loop densities we find
are not greatly influenced by the choice of straight rather than
curved segments in the initial conditions.

\section{Conclusions}

In this work we extend the simulations of Ref.~\cite{Ringeval:2005kr}
to determine the cosmic string loop density at conformal times a
factor of 15 larger than the ones they reached.  We find general
agreement with Ref.~\cite{Ringeval:2005kr} at early times, but after
that there is a different behavior that more closely approximates the
$x^{-2.5}$ (radiation) $x^{-2}$ (matter) behavior expected for any
loop production function that does not rise rapidly at small scales.
It also agrees with the loop densities computed in
Ref.~\cite{Blanco-Pillado:2013qja} starting from the loop production
function.  Given that agreement, it is better to determine the scaling
loop production function and use that to compute the density,
because the loop production function reaches scaling more quickly than
the loop distribution \cite{BlancoPillado:2011dq}.

Both direct determination of loop density from simulation and
extrapolation from the loop production function now give the same
result that the loop density does not diverge faster at small sizes
than $x^{-2.5}$ (radiation) $x^{-2}$ (matter).  Thus there is no
reason to consider loop densities with faster divergences such as
``model 3'' of Ref.~\cite{Abbott:2017mem}.

The difference between $x^{-2.5}$ (radiation) $x^{-2}$ (matter) and
more negative exponents has important observational consequences.
Scaling distributions going as $x^{-2.5}$ (radiation) $x^{-2}$
(matter) may (and in our simulations do) consist primarily of loops
produced at much earlier times.  When gravitational effects are
included, older loops are suppressed, leading to a flat spectrum at
small scales.  But more negative exponents can arise only from
recently produced loops.  Gravitational effects have had less time to
act on such loops, so in such scenarios the loop distribution
continues to grow toward small scales even when gravitation has been
taken into account.  (See Ref.~\cite{Auclair:2019zoz} for a detailed
analysis.)  This is the reason for the stronger bounds on cosmic
strings found by Ref.~\cite{Abbott:2017mem} and different sensitivity
predicted by Ref.~\cite{Auclair:2019wcv} in the case of ``model 3''.
But we see here that this analysis is based on a transient regime that
does not persist in longer simulations.

\section*{Acknowledgments}

We are grateful to Ana Achucarro, Leandros Perivolaropoulos, Tanmay
Vachaspati, and the Lorentz Center in Leiden for organizing the
workshop ``Cosmic Topological Defects: Dynamics and Multi-messenger
Signatures''.  At that workshop we discussed the difficulty of
comparing simulation results when different groups use different
methods.  We agreed with Christophe Ringeval that we would compute the
loop density directly from our simulations while his group would
compute the loop production function from theirs.  This paper fulfills
our side of that agreement.

This work was supported in part by the National Science Foundation
under grant number 1820902. J. J. B.-P. is supported in part by the Spanish 
Ministry MINECO,  MCIU/AEI/FEDER grant (PGC2018-094626-B-C21), the 
Basque Government grant (IT-979-16) and the Basque 
Foundation for Science (IKERBASQUE). 

\bibliography{paper}

\begin{thebibliography}{24}%
\makeatletter
\providecommand \@ifxundefined [1]{%
 \@ifx{#1\undefined}
}%
\providecommand \@ifnum [1]{%
 \ifnum #1\expandafter \@firstoftwo
 \else \expandafter \@secondoftwo
 \fi
}%
\providecommand \@ifx [1]{%
 \ifx #1\expandafter \@firstoftwo
 \else \expandafter \@secondoftwo
 \fi
}%
\providecommand \natexlab [1]{#1}%
\providecommand \enquote  [1]{``#1''}%
\providecommand \bibnamefont  [1]{#1}%
\providecommand \bibfnamefont [1]{#1}%
\providecommand \citenamefont [1]{#1}%
\providecommand \href@noop [0]{\@secondoftwo}%
\providecommand \href [0]{\begingroup \@sanitize@url \@href}%
\providecommand \@href[1]{\@@startlink{#1}\@@href}%
\providecommand \@@href[1]{\endgroup#1\@@endlink}%
\providecommand \@sanitize@url [0]{\catcode `\\12\catcode `\$12\catcode
  `\&12\catcode `\#12\catcode `\^12\catcode `\_12\catcode `\%12\relax}%
\providecommand \@@startlink[1]{}%
\providecommand \@@endlink[0]{}%
\providecommand \url  [0]{\begingroup\@sanitize@url \@url }%
\providecommand \@url [1]{\endgroup\@href {#1}{\urlprefix }}%
\providecommand \urlprefix  [0]{URL }%
\providecommand \Eprint [0]{\href }%
\providecommand \doibase [0]{http://dx.doi.org/}%
\providecommand \selectlanguage [0]{\@gobble}%
\providecommand \bibinfo  [0]{\@secondoftwo}%
\providecommand \bibfield  [0]{\@secondoftwo}%
\providecommand \translation [1]{[#1]}%
\providecommand \BibitemOpen [0]{}%
\providecommand \bibitemStop [0]{}%
\providecommand \bibitemNoStop [0]{.\EOS\space}%
\providecommand \EOS [0]{\spacefactor3000\relax}%
\providecommand \BibitemShut  [1]{\csname bibitem#1\endcsname}%
\let\auto@bib@innerbib\@empty
\bibitem [{\citenamefont {Kibble}(1976)}]{Kibble:1976sj}%
  \BibitemOpen
  \bibfield  {author} {\bibinfo {author} {\bibfnamefont {T.~W.~B.}\
  \bibnamefont {Kibble}},\ }\bibfield  {title} {\enquote {\bibinfo {title}
  {{Topology of Cosmic Domains and Strings}},}\ }\href {\doibase
  10.1088/0305-4470/9/8/029} {\bibfield  {journal} {\bibinfo  {journal} {J.
  Phys.}\ }\textbf {\bibinfo {volume} {A9}},\ \bibinfo {pages} {1387--1398}
  (\bibinfo {year} {1976})}\BibitemShut {NoStop}%
\bibitem [{\citenamefont {Vilenkin}\ and\ \citenamefont
  {Shellard}(2000)}]{Vilenkin:2000jqa}%
  \BibitemOpen
  \bibfield  {author} {\bibinfo {author} {\bibfnamefont {A.}~\bibnamefont
  {Vilenkin}}\ and\ \bibinfo {author} {\bibfnamefont {E.~P.~S.}\ \bibnamefont
  {Shellard}},\ }\href
  {http://www.cambridge.org/mw/academic/subjects/physics/theoretical-physics-and-mathematical-physics/cosmic-strings-and-other-topological-defects?format=PB}
  {\emph {\bibinfo {title} {{Cosmic Strings and Other Topological Defects}}}}\
  (\bibinfo  {publisher} {Cambridge University Press},\ \bibinfo {year}
  {2000})\BibitemShut {NoStop}%
\bibitem [{\citenamefont {Dvali}\ and\ \citenamefont
  {Vilenkin}(2004)}]{Dvali:2003zj}%
  \BibitemOpen
  \bibfield  {author} {\bibinfo {author} {\bibfnamefont {Gia}\ \bibnamefont
  {Dvali}}\ and\ \bibinfo {author} {\bibfnamefont {Alexander}\ \bibnamefont
  {Vilenkin}},\ }\bibfield  {title} {\enquote {\bibinfo {title} {{Formation and
  evolution of cosmic D strings}},}\ }\href {\doibase
  10.1088/1475-7516/2004/03/010} {\bibfield  {journal} {\bibinfo  {journal}
  {JCAP}\ }\textbf {\bibinfo {volume} {0403}},\ \bibinfo {pages} {010}
  (\bibinfo {year} {2004})},\ \Eprint {http://arxiv.org/abs/hep-th/0312007}
  {arXiv:hep-th/0312007 [hep-th]} \BibitemShut {NoStop}%
\bibitem [{\citenamefont {Copeland}\ \emph {et~al.}(2004)\citenamefont
  {Copeland}, \citenamefont {Myers},\ and\ \citenamefont
  {Polchinski}}]{Copeland:2003bj}%
  \BibitemOpen
  \bibfield  {author} {\bibinfo {author} {\bibfnamefont {Edmund~J.}\
  \bibnamefont {Copeland}}, \bibinfo {author} {\bibfnamefont {Robert~C.}\
  \bibnamefont {Myers}}, \ and\ \bibinfo {author} {\bibfnamefont {Joseph}\
  \bibnamefont {Polchinski}},\ }\bibfield  {title} {\enquote {\bibinfo {title}
  {{Cosmic F and D strings}},}\ }\href {\doibase 10.1088/1126-6708/2004/06/013}
  {\bibfield  {journal} {\bibinfo  {journal} {JHEP}\ }\textbf {\bibinfo
  {volume} {06}},\ \bibinfo {pages} {013} (\bibinfo {year} {2004})},\ \Eprint
  {http://arxiv.org/abs/hep-th/0312067} {arXiv:hep-th/0312067 [hep-th]}
  \BibitemShut {NoStop}%
\bibitem [{\citenamefont {Damour}\ and\ \citenamefont
  {Vilenkin}(2005)}]{Damour:2004kw}%
  \BibitemOpen
  \bibfield  {author} {\bibinfo {author} {\bibfnamefont {Thibault}\
  \bibnamefont {Damour}}\ and\ \bibinfo {author} {\bibfnamefont {Alexander}\
  \bibnamefont {Vilenkin}},\ }\bibfield  {title} {\enquote {\bibinfo {title}
  {{Gravitational radiation from cosmic (super)strings: Bursts, stochastic
  background, and observational windows}},}\ }\href {\doibase
  10.1103/PhysRevD.71.063510} {\bibfield  {journal} {\bibinfo  {journal} {Phys.
  Rev.}\ }\textbf {\bibinfo {volume} {D71}},\ \bibinfo {pages} {063510}
  (\bibinfo {year} {2005})},\ \Eprint {http://arxiv.org/abs/hep-th/0410222}
  {arXiv:hep-th/0410222 [hep-th]} \BibitemShut {NoStop}%
\bibitem [{\citenamefont {Siemens}\ \emph {et~al.}(2006)\citenamefont
  {Siemens}, \citenamefont {Creighton}, \citenamefont {Maor}, \citenamefont
  {Ray~Majumder}, \citenamefont {Cannon},\ and\ \citenamefont
  {Read}}]{Siemens:2006vk}%
  \BibitemOpen
  \bibfield  {author} {\bibinfo {author} {\bibfnamefont {Xavier}\ \bibnamefont
  {Siemens}}, \bibinfo {author} {\bibfnamefont {Jolien}\ \bibnamefont
  {Creighton}}, \bibinfo {author} {\bibfnamefont {Irit}\ \bibnamefont {Maor}},
  \bibinfo {author} {\bibfnamefont {Saikat}\ \bibnamefont {Ray~Majumder}},
  \bibinfo {author} {\bibfnamefont {Kipp}\ \bibnamefont {Cannon}}, \ and\
  \bibinfo {author} {\bibfnamefont {Jocelyn}\ \bibnamefont {Read}},\ }\bibfield
   {title} {\enquote {\bibinfo {title} {{Gravitational wave bursts from cosmic
  (super)strings: Quantitative analysis and constraints}},}\ }\href {\doibase
  10.1103/PhysRevD.73.105001} {\bibfield  {journal} {\bibinfo  {journal} {Phys.
  Rev.}\ }\textbf {\bibinfo {volume} {D73}},\ \bibinfo {pages} {105001}
  (\bibinfo {year} {2006})},\ \Eprint {http://arxiv.org/abs/gr-qc/0603115}
  {arXiv:gr-qc/0603115 [gr-qc]} \BibitemShut {NoStop}%
\bibitem [{\citenamefont {Sanidas}\ \emph {et~al.}(2012)\citenamefont
  {Sanidas}, \citenamefont {Battye},\ and\ \citenamefont
  {Stappers}}]{Sanidas:2012ee}%
  \BibitemOpen
  \bibfield  {author} {\bibinfo {author} {\bibfnamefont {S.~A.}\ \bibnamefont
  {Sanidas}}, \bibinfo {author} {\bibfnamefont {R.~A.}\ \bibnamefont {Battye}},
  \ and\ \bibinfo {author} {\bibfnamefont {B.~W.}\ \bibnamefont {Stappers}},\
  }\bibfield  {title} {\enquote {\bibinfo {title} {{Constraints on cosmic
  string tension imposed by the limit on the stochastic gravitational wave
  background from the European Pulsar Timing Array}},}\ }\href {\doibase
  10.1103/PhysRevD.85.122003} {\bibfield  {journal} {\bibinfo  {journal} {Phys.
  Rev.}\ }\textbf {\bibinfo {volume} {D85}},\ \bibinfo {pages} {122003}
  (\bibinfo {year} {2012})},\ \Eprint {http://arxiv.org/abs/1201.2419}
  {arXiv:1201.2419 [astro-ph.CO]} \BibitemShut {NoStop}%
\bibitem [{\citenamefont {Binetruy}\ \emph {et~al.}(2012)\citenamefont
  {Binetruy}, \citenamefont {Bohe}, \citenamefont {Caprini},\ and\
  \citenamefont {Dufaux}}]{Binetruy:2012ze}%
  \BibitemOpen
  \bibfield  {author} {\bibinfo {author} {\bibfnamefont {Pierre}\ \bibnamefont
  {Binetruy}}, \bibinfo {author} {\bibfnamefont {Alejandro}\ \bibnamefont
  {Bohe}}, \bibinfo {author} {\bibfnamefont {Chiara}\ \bibnamefont {Caprini}},
  \ and\ \bibinfo {author} {\bibfnamefont {Jean-Francois}\ \bibnamefont
  {Dufaux}},\ }\bibfield  {title} {\enquote {\bibinfo {title} {{Cosmological
  Backgrounds of Gravitational Waves and eLISA/NGO: Phase Transitions, Cosmic
  Strings and Other Sources}},}\ }\href {\doibase
  10.1088/1475-7516/2012/06/027} {\bibfield  {journal} {\bibinfo  {journal}
  {JCAP}\ }\textbf {\bibinfo {volume} {1206}},\ \bibinfo {pages} {027}
  (\bibinfo {year} {2012})},\ \Eprint {http://arxiv.org/abs/1201.0983}
  {arXiv:1201.0983 [gr-qc]} \BibitemShut {NoStop}%
\bibitem [{\citenamefont {Sousa}\ and\ \citenamefont
  {Avelino}(2016)}]{Sousa:2016ggw}%
  \BibitemOpen
  \bibfield  {author} {\bibinfo {author} {\bibfnamefont {L.}~\bibnamefont
  {Sousa}}\ and\ \bibinfo {author} {\bibfnamefont {P.~P.}\ \bibnamefont
  {Avelino}},\ }\bibfield  {title} {\enquote {\bibinfo {title} {{Probing Cosmic
  Superstrings with Gravitational Waves}},}\ }\href {\doibase
  10.1103/PhysRevD.94.063529} {\bibfield  {journal} {\bibinfo  {journal} {Phys.
  Rev.}\ }\textbf {\bibinfo {volume} {D94}},\ \bibinfo {pages} {063529}
  (\bibinfo {year} {2016})},\ \Eprint {http://arxiv.org/abs/1606.05585}
  {arXiv:1606.05585 [astro-ph.CO]} \BibitemShut {NoStop}%
\bibitem [{\citenamefont {Kuroyanagi}\ \emph {et~al.}(2012)\citenamefont
  {Kuroyanagi}, \citenamefont {Miyamoto}, \citenamefont {Sekiguchi},
  \citenamefont {Takahashi},\ and\ \citenamefont {Silk}}]{Kuroyanagi:2012wm}%
  \BibitemOpen
  \bibfield  {author} {\bibinfo {author} {\bibfnamefont {Sachiko}\ \bibnamefont
  {Kuroyanagi}}, \bibinfo {author} {\bibfnamefont {Koichi}\ \bibnamefont
  {Miyamoto}}, \bibinfo {author} {\bibfnamefont {Toyokazu}\ \bibnamefont
  {Sekiguchi}}, \bibinfo {author} {\bibfnamefont {Keitaro}\ \bibnamefont
  {Takahashi}}, \ and\ \bibinfo {author} {\bibfnamefont {Joseph}\ \bibnamefont
  {Silk}},\ }\bibfield  {title} {\enquote {\bibinfo {title} {{Forecast
  constraints on cosmic string parameters from gravitational wave direct
  detection experiments}},}\ }\href {\doibase 10.1103/PhysRevD.86.023503}
  {\bibfield  {journal} {\bibinfo  {journal} {Phys. Rev.}\ }\textbf {\bibinfo
  {volume} {D86}},\ \bibinfo {pages} {023503} (\bibinfo {year} {2012})},\
  \Eprint {http://arxiv.org/abs/1202.3032} {arXiv:1202.3032 [astro-ph.CO]}
  \BibitemShut {NoStop}%
\bibitem [{\citenamefont {Blanco-Pillado}\ and\ \citenamefont
  {Olum}(2017)}]{Blanco-Pillado:2017oxo}%
  \BibitemOpen
  \bibfield  {author} {\bibinfo {author} {\bibfnamefont {Jose~J.}\ \bibnamefont
  {Blanco-Pillado}}\ and\ \bibinfo {author} {\bibfnamefont {Ken~D.}\
  \bibnamefont {Olum}},\ }\bibfield  {title} {\enquote {\bibinfo {title}
  {{Stochastic gravitational wave background from smoothed cosmic string
  loops}},}\ }\href {\doibase 10.1103/PhysRevD.96.104046} {\bibfield  {journal}
  {\bibinfo  {journal} {Phys. Rev.}\ }\textbf {\bibinfo {volume} {D96}},\
  \bibinfo {pages} {104046} (\bibinfo {year} {2017})},\ \Eprint
  {http://arxiv.org/abs/1709.02693} {arXiv:1709.02693 [astro-ph.CO]}
  \BibitemShut {NoStop}%
\bibitem [{\citenamefont {Abbott}\ \emph {et~al.}(2018)\citenamefont {Abbott}
  \emph {et~al.}}]{Abbott:2017mem}%
  \BibitemOpen
  \bibfield  {author} {\bibinfo {author} {\bibfnamefont {B.~P.}\ \bibnamefont
  {Abbott}} \emph {et~al.} (\bibinfo {collaboration} {LIGO Scientific,
  Virgo}),\ }\bibfield  {title} {\enquote {\bibinfo {title} {{Constraints on
  cosmic strings using data from the first Advanced LIGO observing run}},}\
  }\href {\doibase 10.1103/PhysRevD.97.102002} {\bibfield  {journal} {\bibinfo
  {journal} {Phys. Rev.}\ }\textbf {\bibinfo {volume} {D97}},\ \bibinfo {pages}
  {102002} (\bibinfo {year} {2018})},\ \Eprint
  {http://arxiv.org/abs/1712.01168} {arXiv:1712.01168 [gr-qc]} \BibitemShut
  {NoStop}%
\bibitem [{\citenamefont {Auclair}\ \emph
  {et~al.}(2019{\natexlab{a}})\citenamefont {Auclair} \emph
  {et~al.}}]{Auclair:2019wcv}%
  \BibitemOpen
  \bibfield  {author} {\bibinfo {author} {\bibfnamefont {Pierre}\ \bibnamefont
  {Auclair}} \emph {et~al.},\ }\bibfield  {title} {\enquote {\bibinfo {title}
  {{Probing the gravitational wave background from cosmic strings with
  LISA}},}\ }\href@noop {} {\  (\bibinfo {year} {2019}{\natexlab{a}})},\
  \Eprint {http://arxiv.org/abs/1909.00819} {arXiv:1909.00819 [astro-ph.CO]}
  \BibitemShut {NoStop}%
\bibitem [{\citenamefont {Blanco-Pillado}\ \emph {et~al.}(2018)\citenamefont
  {Blanco-Pillado}, \citenamefont {Olum},\ and\ \citenamefont
  {Siemens}}]{Blanco-Pillado:2017rnf}%
  \BibitemOpen
  \bibfield  {author} {\bibinfo {author} {\bibfnamefont {Jose~J.}\ \bibnamefont
  {Blanco-Pillado}}, \bibinfo {author} {\bibfnamefont {Ken~D.}\ \bibnamefont
  {Olum}}, \ and\ \bibinfo {author} {\bibfnamefont {Xavier}\ \bibnamefont
  {Siemens}},\ }\bibfield  {title} {\enquote {\bibinfo {title} {{New limits on
  cosmic strings from gravitational wave observation}},}\ }\href {\doibase
  10.1016/j.physletb.2018.01.050} {\bibfield  {journal} {\bibinfo  {journal}
  {Phys. Lett.}\ }\textbf {\bibinfo {volume} {B778}},\ \bibinfo {pages}
  {392--396} (\bibinfo {year} {2018})},\ \Eprint
  {http://arxiv.org/abs/1709.02434} {arXiv:1709.02434 [astro-ph.CO]}
  \BibitemShut {NoStop}%
\bibitem [{\citenamefont {Albrecht}\ and\ \citenamefont
  {Turok}(1989)}]{Albrecht:1989mk}%
  \BibitemOpen
  \bibfield  {author} {\bibinfo {author} {\bibfnamefont {Andreas}\ \bibnamefont
  {Albrecht}}\ and\ \bibinfo {author} {\bibfnamefont {Neil}\ \bibnamefont
  {Turok}},\ }\bibfield  {title} {\enquote {\bibinfo {title} {{Evolution of
  Cosmic String Networks}},}\ }\href {\doibase 10.1103/PhysRevD.40.973}
  {\bibfield  {journal} {\bibinfo  {journal} {Phys. Rev.}\ }\textbf {\bibinfo
  {volume} {D40}},\ \bibinfo {pages} {973--1001} (\bibinfo {year}
  {1989})}\BibitemShut {NoStop}%
\bibitem [{\citenamefont {Ringeval}\ \emph {et~al.}(2007)\citenamefont
  {Ringeval}, \citenamefont {Sakellariadou},\ and\ \citenamefont
  {Bouchet}}]{Ringeval:2005kr}%
  \BibitemOpen
  \bibfield  {author} {\bibinfo {author} {\bibfnamefont {Christophe}\
  \bibnamefont {Ringeval}}, \bibinfo {author} {\bibfnamefont {Mairi}\
  \bibnamefont {Sakellariadou}}, \ and\ \bibinfo {author} {\bibfnamefont
  {Francois}\ \bibnamefont {Bouchet}},\ }\bibfield  {title} {\enquote {\bibinfo
  {title} {{Cosmological evolution of cosmic string loops}},}\ }\href {\doibase
  10.1088/1475-7516/2007/02/023} {\bibfield  {journal} {\bibinfo  {journal}
  {JCAP}\ }\textbf {\bibinfo {volume} {0702}},\ \bibinfo {pages} {023}
  (\bibinfo {year} {2007})},\ \Eprint {http://arxiv.org/abs/astro-ph/0511646}
  {arXiv:astro-ph/0511646 [astro-ph]} \BibitemShut {NoStop}%
\bibitem [{\citenamefont {Vanchurin}\ \emph {et~al.}(2006)\citenamefont
  {Vanchurin}, \citenamefont {Olum},\ and\ \citenamefont
  {Vilenkin}}]{Vanchurin:2005pa}%
  \BibitemOpen
  \bibfield  {author} {\bibinfo {author} {\bibfnamefont {Vitaly}\ \bibnamefont
  {Vanchurin}}, \bibinfo {author} {\bibfnamefont {Ken~D.}\ \bibnamefont
  {Olum}}, \ and\ \bibinfo {author} {\bibfnamefont {Alexander}\ \bibnamefont
  {Vilenkin}},\ }\bibfield  {title} {\enquote {\bibinfo {title} {{Scaling of
  cosmic string loops}},}\ }\href {\doibase 10.1103/PhysRevD.74.063527}
  {\bibfield  {journal} {\bibinfo  {journal} {Phys. Rev.}\ }\textbf {\bibinfo
  {volume} {D74}},\ \bibinfo {pages} {063527} (\bibinfo {year} {2006})},\
  \Eprint {http://arxiv.org/abs/gr-qc/0511159} {arXiv:gr-qc/0511159 [gr-qc]}
  \BibitemShut {NoStop}%
\bibitem [{\citenamefont {Olum}\ and\ \citenamefont
  {Vanchurin}(2007)}]{Olum:2006ix}%
  \BibitemOpen
  \bibfield  {author} {\bibinfo {author} {\bibfnamefont {Ken~D.}\ \bibnamefont
  {Olum}}\ and\ \bibinfo {author} {\bibfnamefont {Vitaly}\ \bibnamefont
  {Vanchurin}},\ }\bibfield  {title} {\enquote {\bibinfo {title} {{Cosmic
  string loops in the expanding Universe}},}\ }\href {\doibase
  10.1103/PhysRevD.75.063521} {\bibfield  {journal} {\bibinfo  {journal} {Phys.
  Rev.}\ }\textbf {\bibinfo {volume} {D75}},\ \bibinfo {pages} {063521}
  (\bibinfo {year} {2007})},\ \Eprint {http://arxiv.org/abs/astro-ph/0610419}
  {arXiv:astro-ph/0610419 [astro-ph]} \BibitemShut {NoStop}%
\bibitem [{\citenamefont {Blanco-Pillado}\ \emph {et~al.}(2011)\citenamefont
  {Blanco-Pillado}, \citenamefont {Olum},\ and\ \citenamefont
  {Shlaer}}]{BlancoPillado:2011dq}%
  \BibitemOpen
  \bibfield  {author} {\bibinfo {author} {\bibfnamefont {Jose~J.}\ \bibnamefont
  {Blanco-Pillado}}, \bibinfo {author} {\bibfnamefont {Ken~D.}\ \bibnamefont
  {Olum}}, \ and\ \bibinfo {author} {\bibfnamefont {Benjamin}\ \bibnamefont
  {Shlaer}},\ }\bibfield  {title} {\enquote {\bibinfo {title} {{Large parallel
  cosmic string simulations: New results on loop production}},}\ }\href
  {\doibase 10.1103/PhysRevD.83.083514} {\bibfield  {journal} {\bibinfo
  {journal} {Phys. Rev.}\ }\textbf {\bibinfo {volume} {D83}},\ \bibinfo {pages}
  {083514} (\bibinfo {year} {2011})},\ \Eprint {http://arxiv.org/abs/1101.5173}
  {arXiv:1101.5173 [astro-ph.CO]} \BibitemShut {NoStop}%
\bibitem [{\citenamefont {Blanco-Pillado}\ \emph {et~al.}(2014)\citenamefont
  {Blanco-Pillado}, \citenamefont {Olum},\ and\ \citenamefont
  {Shlaer}}]{Blanco-Pillado:2013qja}%
  \BibitemOpen
  \bibfield  {author} {\bibinfo {author} {\bibfnamefont {Jose~J.}\ \bibnamefont
  {Blanco-Pillado}}, \bibinfo {author} {\bibfnamefont {Ken~D.}\ \bibnamefont
  {Olum}}, \ and\ \bibinfo {author} {\bibfnamefont {Benjamin}\ \bibnamefont
  {Shlaer}},\ }\bibfield  {title} {\enquote {\bibinfo {title} {{The number of
  cosmic string loops}},}\ }\href {\doibase 10.1103/PhysRevD.89.023512}
  {\bibfield  {journal} {\bibinfo  {journal} {Phys. Rev.}\ }\textbf {\bibinfo
  {volume} {D89}},\ \bibinfo {pages} {023512} (\bibinfo {year} {2014})},\
  \Eprint {http://arxiv.org/abs/1309.6637} {arXiv:1309.6637 [astro-ph.CO]}
  \BibitemShut {NoStop}%
\bibitem [{\citenamefont {Blanco-Pillado}\ \emph {et~al.}(2019)\citenamefont
  {Blanco-Pillado}, \citenamefont {Olum},\ and\ \citenamefont
  {Wachter}}]{Blanco-Pillado:2019vcs}%
  \BibitemOpen
  \bibfield  {author} {\bibinfo {author} {\bibfnamefont {Jose~J.}\ \bibnamefont
  {Blanco-Pillado}}, \bibinfo {author} {\bibfnamefont {Ken~D.}\ \bibnamefont
  {Olum}}, \ and\ \bibinfo {author} {\bibfnamefont {Jeremy~M.}\ \bibnamefont
  {Wachter}},\ }\bibfield  {title} {\enquote {\bibinfo {title}
  {{Energy-conservation constraints on cosmic string loop production and
  distribution functions}},}\ }\href {\doibase 10.1103/PhysRevD.100.123526}
  {\bibfield  {journal} {\bibinfo  {journal} {Phys. Rev.}\ }\textbf {\bibinfo
  {volume} {D100}},\ \bibinfo {pages} {123526} (\bibinfo {year} {2019})},\
  \Eprint {http://arxiv.org/abs/1907.09373} {arXiv:1907.09373 [astro-ph.CO]}
  \BibitemShut {NoStop}%
\bibitem [{\citenamefont {Vachaspati}\ and\ \citenamefont
  {Vilenkin}(1984)}]{Vachaspati:1984dz}%
  \BibitemOpen
  \bibfield  {author} {\bibinfo {author} {\bibfnamefont {Tanmay}\ \bibnamefont
  {Vachaspati}}\ and\ \bibinfo {author} {\bibfnamefont {Alexander}\
  \bibnamefont {Vilenkin}},\ }\bibfield  {title} {\enquote {\bibinfo {title}
  {{Formation and Evolution of Cosmic Strings}},}\ }\href {\doibase
  10.1103/PhysRevD.30.2036} {\bibfield  {journal} {\bibinfo  {journal} {Phys.
  Rev.}\ }\textbf {\bibinfo {volume} {D30}},\ \bibinfo {pages} {2036} (\bibinfo
  {year} {1984})}\BibitemShut {NoStop}%
\bibitem [{\citenamefont {Lorenz}\ \emph {et~al.}(2010)\citenamefont {Lorenz},
  \citenamefont {Ringeval},\ and\ \citenamefont
  {Sakellariadou}}]{Lorenz:2010sm}%
  \BibitemOpen
  \bibfield  {author} {\bibinfo {author} {\bibfnamefont {Larissa}\ \bibnamefont
  {Lorenz}}, \bibinfo {author} {\bibfnamefont {Christophe}\ \bibnamefont
  {Ringeval}}, \ and\ \bibinfo {author} {\bibfnamefont {Mairi}\ \bibnamefont
  {Sakellariadou}},\ }\bibfield  {title} {\enquote {\bibinfo {title} {{Cosmic
  string loop distribution on all length scales and at any redshift}},}\ }\href
  {\doibase 10.1088/1475-7516/2010/10/003} {\bibfield  {journal} {\bibinfo
  {journal} {JCAP}\ }\textbf {\bibinfo {volume} {1010}},\ \bibinfo {pages}
  {003} (\bibinfo {year} {2010})},\ \Eprint {http://arxiv.org/abs/1006.0931}
  {arXiv:1006.0931 [astro-ph.CO]} \BibitemShut {NoStop}%
\bibitem [{\citenamefont {Auclair}\ \emph
  {et~al.}(2019{\natexlab{b}})\citenamefont {Auclair}, \citenamefont
  {Ringeval}, \citenamefont {Sakellariadou},\ and\ \citenamefont
  {Steer}}]{Auclair:2019zoz}%
  \BibitemOpen
  \bibfield  {author} {\bibinfo {author} {\bibfnamefont {Pierre}\ \bibnamefont
  {Auclair}}, \bibinfo {author} {\bibfnamefont {Christophe}\ \bibnamefont
  {Ringeval}}, \bibinfo {author} {\bibfnamefont {Mairi}\ \bibnamefont
  {Sakellariadou}}, \ and\ \bibinfo {author} {\bibfnamefont {Daniele}\
  \bibnamefont {Steer}},\ }\bibfield  {title} {\enquote {\bibinfo {title}
  {{Cosmic string loop production functions}},}\ }\href {\doibase
  10.1088/1475-7516/2019/06/015} {\bibfield  {journal} {\bibinfo  {journal}
  {JCAP}\ }\textbf {\bibinfo {volume} {1906}},\ \bibinfo {pages} {015}
  (\bibinfo {year} {2019}{\natexlab{b}})},\ \Eprint
  {http://arxiv.org/abs/1903.06685} {arXiv:1903.06685 [astro-ph.CO]}
  \BibitemShut {NoStop}%
\end{thebibliography}%

\end{document}